\documentclass[aps,
amsmath,amssymb,
pre,reprint,showpacs,superscriptaddress]{revtex4-1}

\usepackage{graphicx}
\usepackage{dcolumn}
\usepackage{bm}

\begin{document}


\title{Automodel solutions for L\'evy flight-based transport on a uniform background}


\author{A. B. Kukushkin}
\email[E-mail address: ]{kukushkin\_ab@nrcki.ru}
\affiliation{National Research Center ''Kurchatov Institute'', Moscow, 123182, Russian Federation}
\affiliation{National Research Nuclear University MEPhI, Moscow, 115409, Russian Federation}

\author{P. A. Sdvizhenskii}
\affiliation{National Research Center ''Kurchatov Institute'', Moscow, 123182, Russian Federation}


\date{\today}

\begin{abstract}
A wide class of non-stationary superdiffusive transport on a uniform background with a power-law decay, at large distances, 
of the step-length probability distribution function (PDF) is shown to possess an automodel solution. 
The solution for Green function is constructed using the scaling laws for the propagation front 
(relevant-to-superdiffusion average displacement) and asymptotic solutions far beyond 
and far in advance of the propagation front. These scaling laws are determined essentially 
by the long-free-path carriers (L\'evy flights). The validity of the suggested automodel solution is proved 
by its comparison with numerical solutions in the one-dimensional (1D) case of the transport equation 
with a simple long-tailed PDF with various power-law exponents and in the 3D case of 
the Biberman-Holstein equation of the resonance radiation transfer for various 
(Doppler, Lorentz, Voight and Holtsmark) spectral line shapes.
\end{abstract}

\pacs{05.40.Fb}

\maketitle

\section{\label{sec1:intro}Introduction}

Self-similarity of random walks is expressed in the fact that the space-time evolution of the transport on a uniform background 
from an instant point source (i.e.\ Green function) is a function 
of a single variable. In the case of normal (or ordinary) diffusion, defined as the Brownian motion 
described by the differential equation of the Fokker-Planck type, the Green function is a Gaussian 
which argument determines the scaling law for the propagation front, $r_{\text{fr}} \sim (Dt)^{1/2}$,
where $D$ is the diffusion coefficient.

In a broader frame, beyond the Brownian motion, the diversity of the free-path lengths (step lengths) 
provided by the long-tailed, power-law probability distribution function (PDF) leads to a superdiffusive transport 
(see, e.g., \cite{shleza:ny,duspuch:bif}). In this case the dominant contribution to the transport comes from 
the long-free-path carriers (named, by P. Mandelbrot \cite{mandelbrot}, L\'evy flights, 
cf.\ page IX in Ref.\ \cite{shleza:ny}). In various physics problems, 
superdiffusion was qualified/named as a non-local transport which is described by an integral, 
in space variables, equation which is non-reducible to a differential one.

This is the case, e.g., for the Biberman-Holstein equation \cite{bib:jetp,holstein:physrev} which 
is derived from a couple of differential kinetic equation for photons and atoms, or ions, 
and describes the space-time evolution of the density of excited atoms/ions produced 
by the radiative transfer in the spectral lines in gases and plasmas. The Biberman-Holstein equation assumes 
complete redistribution over frequency (within spectral line width) in the elementary act of the resonance 
scattering (i.e.\ absorption and subsequent emission) of the photon by an atom/ion. It is the loss of memory 
by the photon during its trapping by the atom/ion that makes the process a Markovian one. The respective value 
of $D$ in the differential equation, derived by a proper expansion/simplification 
of the integral equation, explicitly depends on the size, $L$, of a finite medium 
and tends to infinity with $L \to \infty$ (see, e.g., \cite{bibvoyak:noneq,abkolis:vtp1987}). The latter makes 
the very concept of diffusivity irrelevant to such a mechanism of transport (however, the term diffusion 
is sometimes applied to such phenomena). The respective mean squared displacement 
in an infinite medium also diverges so that the propagation front should be defined 
in a way relevant to superdiffusion \cite{vekl:jetp} (see also \cite{bibvoyak:noneq,abkolis:vtp1987}). The role 
of long-free-path photons, i.e.\ those emitted in the far wings of the spectral line shape, 
has been recognized \cite{bib:dan,kogan:iaea1968} and was the basis 
of the escape probability (EP) approaches \cite{radtransf:cambr,napart:teplo}. We can also mention 
an extension of the Biberman-Holstein approach from the line to continuous spectra: 
the EP method for non-stationary heat transport by the longitudinal (electron Bernstein) 
waves in plasmas \cite{kuklissav:jetp}, which extended the integral equation approach to the similar 
steady-state transport \cite{rosliu:physfluids}, and the EP approach to the transport by plasma waves \cite{kuk:jetplett}, 
which extended/modified the approach to the transport by electron cyclotron waves 
in nuclear fusion plasmas \cite{tamor:fustech}. Despite the term ``L\'evy flights'' was not used in Refs.\ from 
\cite{bib:jetp} to \cite{tamor:fustech}, the dominant mechanism of the transport considered 
in these papers actually gets to this category.

Alternatively to the Biberman-Holstein equation widely used in the laboratory plasmas, in astrophysics 
the above-mentioned couple of differential kinetic equations for photons and atoms/ions 
is often reduced to an integral, in space variables, equation for the radiation intensity 
(cf., e.g., \cite{ivanov:radtransf,mih:stellaratm}). Here the role of L\'evy flights is identified, as a rule without naming, as well.

Now situation in the literature appears to be ambiguous. On the one hand, in the article
 ``A L\'evy flight for light'' \cite{barberwier:nature} it was claimed that ``\dots\ to date, it has not seemed possible 
 to observe and study L\'evy transport in actual materials. For example, experimental work on heat, 
 sound, and light diffusion is generally limited to normal, brownian, diffusion.'' 
 On the other hand, the role of L\'evy flight for light in the conventional frames 
 of radiative transfer in spectral lines is identified, e.g., in Ref.\ \cite{merchegue:pra}.

In the recent decades, the original version of superdiffusion in the formalism of fractional order derivatives
 went far beyond the pure mathematics and now is widely used in many applications 
 (see, e.g., \cite{shleza:ny,duspuch:bif,uchai:physusp}). In this formalism the scaling laws 
 (i.e.\ self-similarity of solutions) are derived much easier than in the case of integral equations, 
 however derivation of an exact solution is often as difficult as in the integral equation formalisms.

In the present paper we formulate a method for derivation of automodel solutions for a wide 
class of non-stationary superdiffusive transport on a uniform background in the frame of integral 
equation formalism. The method allows derivation of the solution in the form which relies on 
the scaling law for the propagation front, defined as a relevant-to-superdiffusion average displacement. 
The algorithm of constructing such a solution uses also the asymptotic behavior of the Green function 
far beyond and far in advance of the propagation front. All the building blocks of the model are shown 
to be determined by the long-free-path carriers in the power-law tail of the PDF (i.e., by L\'evy flights). 
The validity of the suggested automodel solution is proved by its comparison with numerical solutions 
of transport equations. First, the main principles of the method are presented for the one-dimensional 
(1D) transport with a simple long-tailed PDF (Sec.\ 2). A test of the proposed automodel solution 
is carried out via comparison with direct numerical solution for various power-law exponents (Sec.\ 3). 
A 3D transport is addressed in the case of the Biberman-Holstein equation with various spectral 
line shapes (Sec.\ 4). A test of the general automodel solution is made via comparison with analytic 
solutions by Veklenko \cite{vekl:jetp} for the Green function for the Doppler and Lorenz spectral line shapes, 
and with similar solutions for the Voight and Holtsmark spectral line shapes.

\section{\label{sec2:method}A method of deriving an automodel Green function (simple PDF, 1D case)}

We consider the 1D transport on a uniform background, described by the equation for spatial density $f(x,t)$ 
of an excitation of the background medium, which may evolve due to the exchange of excitation between 
various points of the medium via emission and absorption of the carriers (here the retardation 
caused by the finite velocity of carriers is neglected):
\begin{multline}
\frac{\partial f(x,t)}{\partial t} = \frac{1}{\tau} \int_{-\infty}^{\infty} 
W\bigl( \left\vert x - x_1 \right\vert \bigr)
f(x_1,t)\,dx_1 \\
- \left( \frac{1}{\tau} + \sigma \right) f(x,t) + q(x,t),
\label{eq:transp}
\end{multline}
where $W(x)$ is a step-length PDF (i.e.\ the probability that the carrier, emitted at some point, 
is absorbed at a distance $x$ from that point), $1/\tau$ is the absolute value of the emission rate, 
$q$ is the source function, which is the rate of production of excitation by an external source 
(i.e.\ a source which differs from the excitation of the medium due to absorption described by the $W$ function), 
and $\sigma$ is the rate of quenching of excitation. The uniformity of background assumes that, 
first, the $W$ is a function of only one variable, the distance between the points of emission and 
absorption, and, second, $\tau$ and $\sigma$ are the constants. The latter makes the role of quenching simply 
described by the time exponent $\exp(-\sigma t)$, therefore in what follows we omit this process. 
Hereafter we use dimensionless time and space coordinate, assuming the normalization of time by 
$\tau$ and using a dimensionless PDF. We will seek for the Green function, taking, respectively, 
the source function as a point instant source,
\begin{equation}
\label{eq:source}
q(x,t) = \delta(x)\delta(t).
\end{equation}
We take the PDF in the following simple form which possesses a long-tail and the infinite value of the 
mean square displacement:
\begin{gather}
W(\rho) = \frac{\gamma}{2(1 + \rho)^{\gamma + 1}}, \quad 0 < \gamma < 2, \quad \rho = | x - x' |, \nonumber \\
\int_{-\infty}^{\infty} W\bigl( |x - x'| \bigr)\,dx' = 1.
\label{eq:pdf}
\end{gather}
It is worth to introduce the probability, $T(\rho)$, for the carrier to pass, without any absorption, the distance not exceeding 
a certain value, $\rho$. This function may be expressed in terms of the PDF of Eq.\ (\ref{eq:pdf}):
\begin{equation}
\label{eq:pdfholst}
W(\rho) = - \frac{dT(\rho)}{2d\rho}, \quad T(0) = 1.
\end{equation}
Further, we suggest the following equation for the propagation front, $\rho_{\text{fr}}(t)$, 
which, as shown below, appears to be close to the time dependence of the mean displacement:
\begin{equation}
\label{eq:frontdef}
(t + 1)\,T(\rho_{\text{fr}}(t)) = 1, \quad \rho = |x|.
\end{equation}
Equation (\ref{eq:frontdef}) may be obviously considered as a definition of the time evolution 
of the front, $t_{\text{fr}}(\rho)$. Note that the Eq.\ (\ref{eq:frontdef}) may be substantiated for large 
values of dimensionless time, whereas for $t \sim 1$ it is interpolated to an obvious condition $\rho_{\text{fr}}(0) = 0$.

Equation (\ref{eq:frontdef}) is suggested by the success of the escape probability methods in the theory 
of radiative transfer in spectral lines. These methods have been anticipated by the approximate 
solution \cite{bib:dan}, obtained for steady-state problem by taking $f(x)$ out of the integral term, 
and by the analysis \cite{kogan:iaea1968} of validity of solution \cite{bib:dan}.

Equation (\ref{eq:transp}) allows derivation of asymptotic behavior of the Green function 
far beyond and far in advance of the propagation front defined by Eq.\ (\ref{eq:frontdef}). 
Far in advance of the propagation front arrival at a distance $\rho$, 
$\rho \gg \rho_{\text{fr}}(t) \gg 1$ 
(or, equivalently, for a short time, $1 \ll t \ll t_{\text{fr}}(\rho)$), the density is determined 
by the direct population by the carriers emitted by the source. At this stage the distant 
points see almost a point source of carriers because the diffusion of the excitation around the origin 
$(x=0)$ may be neglected. The exchange of the neighboring distant points may also be neglected. 
This corresponds to the excitation of the distant points by the rare long-free-path 
carriers (i.e. L\'evy flights) whereas the frequent exchange with short-free-path carriers, 
both around the source and the every distant point, may be neglected. This gives a simple relation
\begin{equation}
\label{eq:asympt1}
f(x,t) \approx t\,W(\rho), \quad \rho = |x|, 
\quad \rho \gg \rho_{\text{fr}}(t) \gg 1.
\end{equation}
The asymptotics of the Green function far behind the propagation front, 
$\rho \ll \rho_{\text{fr}}(t)$, or equivalently $t \gg t_{\text{fr}}(\rho) \gg 1$, 
may be found taking into account the above-mentioned frequent exchange with 
short-free-path carriers. The latter produces local uniformity of the density. 
Assuming a plateau-like spatial distribution around the origin, one has:
\begin{equation}
\label{eq:asymptplat}
f(x,t) \sim \frac{1}{2\rho_{\text{fr}}(t)} \eta \bigl( \rho_{\text{fr}}(t) - |x| \bigr), \quad \rho \ll \rho_{\text{fr}}(t).
\end{equation}
where $\eta$ is the Heaviside step function.

It appears that the asymptotic laws of Eq.\ (\ref{eq:asympt1}) and Eq.\ (\ref{eq:asymptplat}) may be unified in a single 
interpolation formula, e.g., of the following type:
\begin{equation}
\label{eq:interp1}
f(x,t) \sim t\,W \! \left( \! \sqrt{x^2 + C \rho_{\text{fr}}^2(t)} \, \right),
\end{equation}
where the constant $C$ should give equality to Eq.\ (\ref{eq:asymptplat}). It's worth, however, 
to leave a functional freedom of the interpolation between Eq.\ (\ref{eq:asympt1}) and Eq.\ (\ref{eq:asymptplat}). This gives
\begin{equation}
\label{eq:interp2}
f(x,t) = t\,W \! \left( \rho g \! \left(\frac{\rho_{\text{fr}}(t)}{\rho}\right) \! \right), \quad \rho = |x|,
\end{equation}
where $g$ is a function of a single variable, and its asymptotic behavior is known:
\begin{gather}
\label{eq:g1} g(s) = 1, \quad s \ll 1,\\
\label{eq:g2} g(s) \propto s, \quad s \gg 1,
\end{gather}
A test of the solution (\ref{eq:interp2})-(\ref{eq:g2}), (\ref{eq:frontdef}) via comparison with 
direct numerical solution of Eq.\ (\ref{eq:transp}) may be considered as an inverse problem 
of reconstructing the function $g$ or, equivalently, as a proof of the scaling law 
(self-similarity of solutions) formulated with Eqs.\ (\ref{eq:interp2})--(\ref{eq:g2}), (\ref{eq:frontdef}).

\section{\label{sec3:model}Validation and reconstruction of automodel Green function from numerical solution 
(simple PDF, 1D case)}

We try the validity of the method of Sec.\ \ref{sec2:method} in the case of a simple PDF with power-law tail. 
We consider the following PDF:
\begin{align}
T(\rho) = \frac{1}{(1 + \rho)^{\gamma}}, &\quad W(\rho) = 
\frac{\gamma}{2(1 + \rho)^{\gamma + 1}}, \nonumber \\
\gamma > 0, &\quad \rho = |x - x'|.
\label{eq:model}
\end{align}
Equation (\ref{eq:transp}) may be solved via Fourier transformation. For the Fourier transform
\begin{equation}
\label{eq:fourier1}
\hat y(p) = \dfrac{1}{\sqrt{2\pi}} \int_{-\infty}^{\infty} y(x) e^{-ipx} \, dx
\end{equation}
with allowance for the relation
\begin{equation}
\label{eq:fourier2}
\widehat{(y*z)} = \sqrt{2\pi} \, \hat y \hat z,
\end{equation}
where
\begin{equation*}
(y*z)(x) = 
\int_{-\infty}^{\infty} y(x-x')z(x') \, dx',
\end{equation*}
equation (\ref{eq:transp}) turns into
\begin{equation}
\label{eq:transpfourier}
\dfrac{\partial \hat f(p,t)}{\partial t} = \sqrt{2\pi} \, \hat W(p) \hat f(p,t) - \hat f(p,t),
\end{equation}
with the boundary condition $\hat f(p,0) = 1/\sqrt{2\pi}$. Solving the Eq.\ (\ref{eq:transpfourier}) and taking the inverse Fourier transform, one has:
\begin{equation}
\label{eq:transpsol}
f(x,t) = \frac{1}{2\pi} e^{-t} \int_{-\infty}^{\infty} e^{\sqrt{2\pi}\, t \hat W(p)} e^{ipx} \, dp,
\end{equation}
where
\begin{align}
\hat W(p) & \equiv \dfrac{1}{\sqrt{2\pi}} \int_{-\infty}^{\infty} 
W(x) e^{-ipx} \, dx \nonumber \\
& = \dfrac{1}{\sqrt{2\pi}} \left\{1 - p\int_{0}^{\infty} \! \dfrac{\sin px}{(1+x)^{\gamma}} \, dx \right\}.
\label{eq:corefourier}
\end{align}
This finally leads to an analytic solution of Eq. (\ref{eq:transp}) with the PDF of Eq. (\ref{eq:model}):
\begin{equation}
\label{eq:transpsolfin}
f(x,t) = \frac{1}{2\pi} \int_{-\infty}^{\infty} \cos px \,
\exp \left( -tp\int_{0}^{\infty} \! \dfrac{\sin px}{(1+x)^{\gamma}} \, dx \right) \, dp.
\end{equation}
In the case of $t \to \infty, x \to 0, 0 < \gamma < 2$ the Eq.\ (\ref{eq:transpsolfin}) 
takes the form
\begin{equation}
f(x,t) = \dfrac{\Gamma(1 + 1/\gamma)}{\pi [t\, I(\gamma)]^{1/\gamma}}.
\label{eq:asymptofexact}
\end{equation}
The respective automodel solution obtained from Eq. (\ref{eq:interp2}) is as follows:
\begin{equation}
\label{eq:solauto}
f(x,t) = t \, \dfrac{\gamma}{2 \left[1 + \rho g \left( \dfrac{\rho_{\text{fr}}(t)}{\rho} \right) \right]^{\gamma+1}}, \quad
\rho = |x|,
\end{equation}
where asymptotics of the function $g$ obeys Eqs.\ (\ref{eq:g1}), (\ref{eq:g2}), and
\begin{equation}
\label{eq:frontmodel}
\rho_{\text{fr}}(t) = (t + 1)^{1/\gamma} - 1.
\end{equation}
In the case of PDF (\ref{eq:model}), Eq.\ (\ref{eq:g2}) may be specified:
\begin{gather}
g(s) = \alpha s, \quad s \gg 1, \nonumber \\
\alpha(\gamma) = \left\{ \dfrac{\pi}{2} \dfrac{\gamma \left[ I(\gamma) \right]^{1/\gamma}}
{\Gamma(1 + 1/\gamma)} \right\}^{1/(\gamma + 1)}, \quad
I(\gamma) = \int_0^{\infty} \dfrac{\sin x}{x^{\gamma}} \, dx.
\label{eq:autofunasympt}
\end{gather}
It appears that the asymptotic solution far behind the front deviates from the qualitative scaling law 
of Eq.\ (\ref{eq:asymptplat}) by the constant, a factor $2\Gamma(1+1/\gamma)/[\pi (I(\gamma))^{1/\gamma}]$ which is of the order of unity, 
$I(1/2) \approx 1.25$, $I(1) \approx 1.56$, $I(3/2) \approx 2.31$.

\begin{figure}
(a)˜\includegraphics[width=0.45\textwidth]{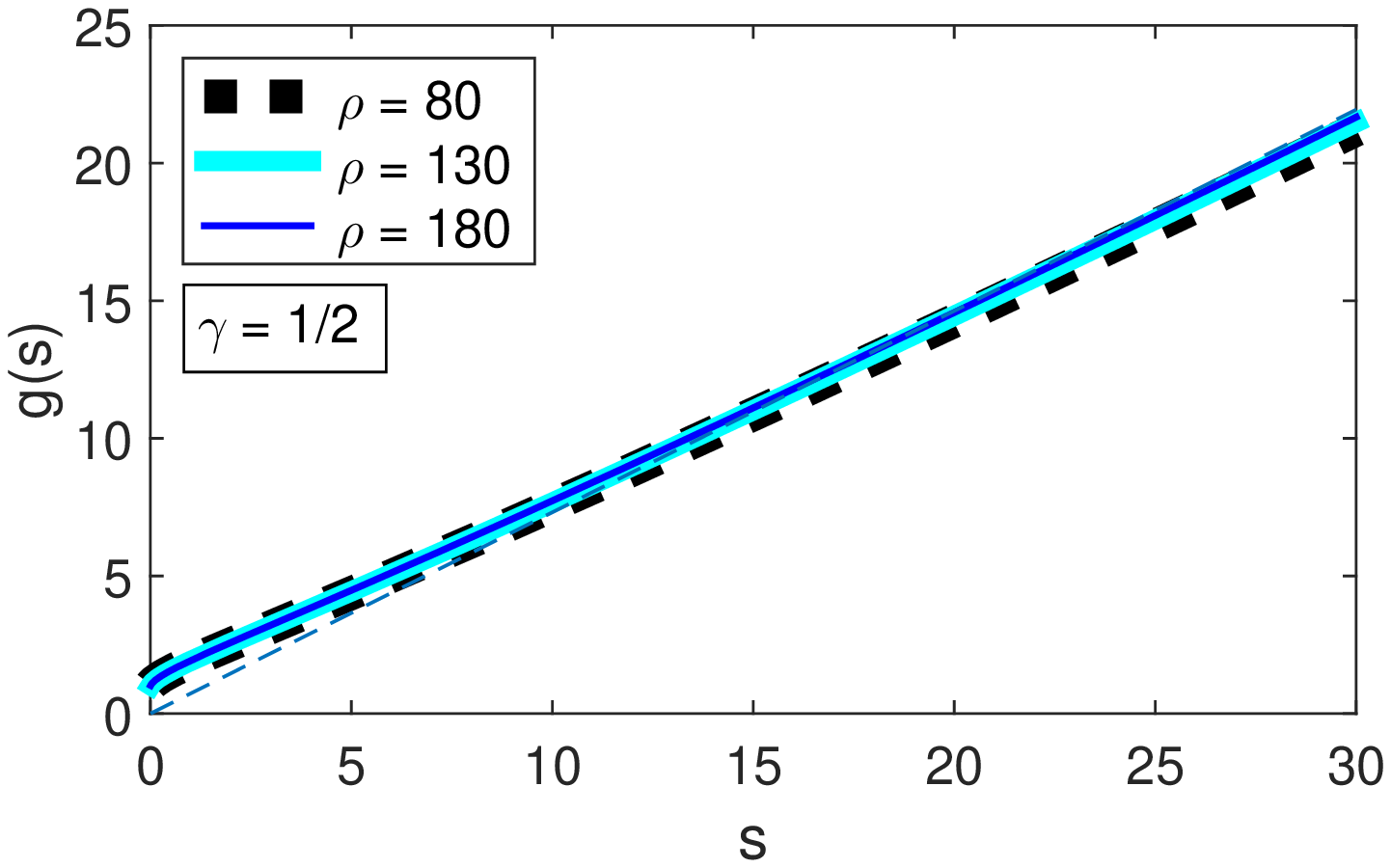} \\
(b)˜\includegraphics[width=0.45\textwidth]{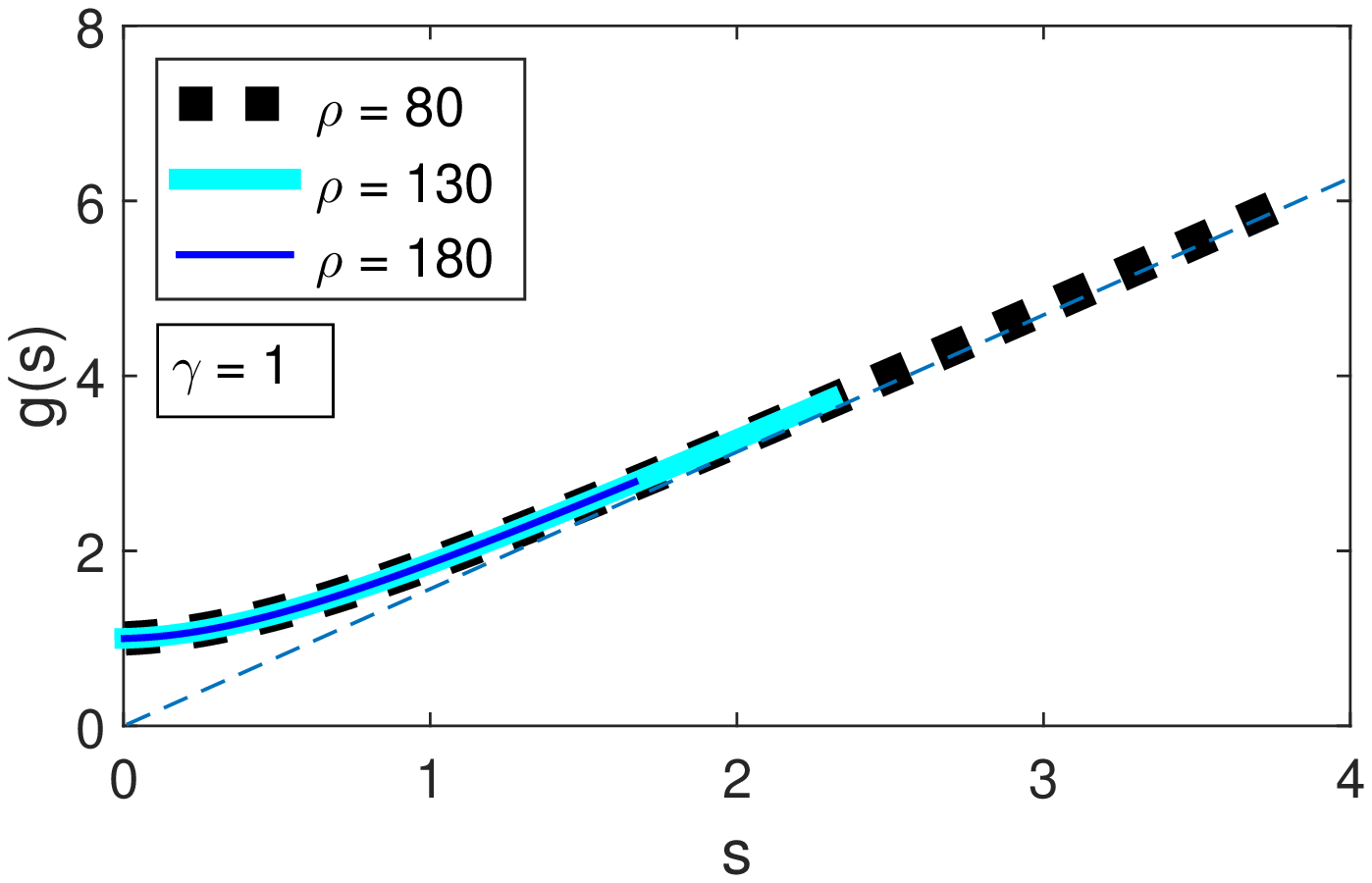} \\
(c)˜\includegraphics[width=0.45\textwidth]{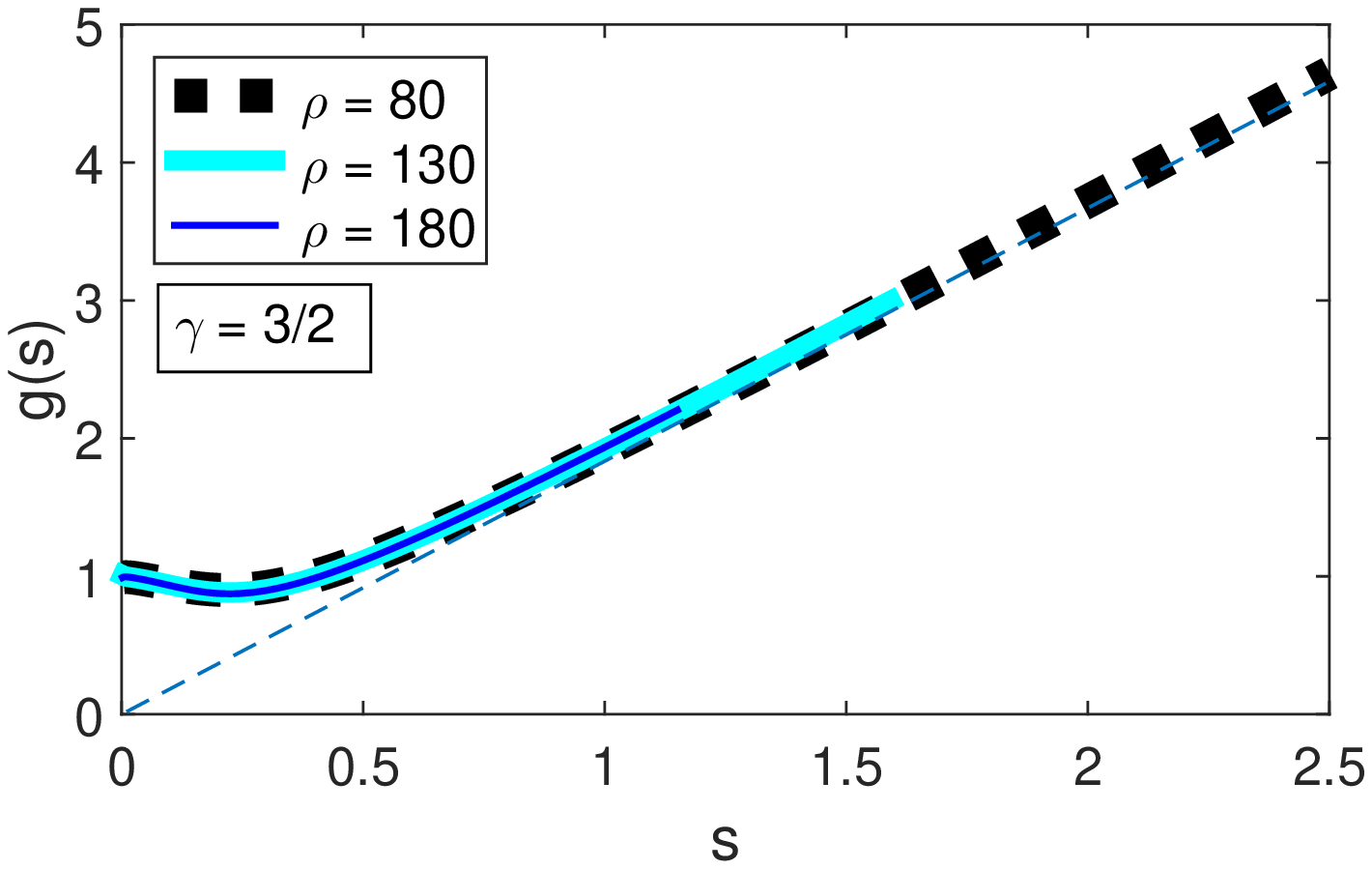}
\caption{\label{fig1} (Color online) The reconstruction of the function $g(s)$, 
$s \equiv \rho_{\text{fr}}(t)/\rho$, via comparing 
Eqs.\ (\ref{eq:solauto}), (\ref{eq:frontmodel}) with computation of 
Eq.\ (\ref{eq:transpsolfin}) for various values of $\gamma$: 
(a) $\gamma = 1/2$; (b) $\gamma = 1$; (c) $\gamma = 3/2$. 
The thin blue dashed line shows the asymptotics (\ref{eq:autofunasympt}) of 
the function $g(s)$.}
\end{figure}

Comparison of Eqs.\ (\ref{eq:solauto}), (\ref{eq:frontmodel}) 
with numerical calculation of Eq.\ (\ref{eq:transpsolfin}) 
enables us, first, to prove the validity of the automodel solution (\ref{eq:solauto}), (\ref{eq:frontmodel}) 
and, second, reconstruct the function $g$. The results of comparison for 
$\gamma=1/2$, $1$ and $3/2$ are presented in Fig.\ \ref{fig1}.

It is seen that for given values of $\gamma$ the function (\ref{eq:solauto}) is indeed an automodel 
solution of Eq.\ (\ref{eq:transp}) with the PDF of Eq.\ (\ref{eq:model}).

\section{\label{sec:autoBHsol}Automodel solution of Biberman-Holstein equation}

The Biberman-Holstein equation for radiative transfer in a uniform medium of two-level 
atoms/ions is obtained from a system of equations for spatial density of excited atoms, 
 $F({\bf r},t)$, and spectral intensity of resonance radiation. This system is reduced to 
 a single equation for $F({\bf r},t)$, which appears to be an integral equation, non-reducible to a differential 
 diffusion-type equation :
\begin{multline}
\dfrac{\partial F({\bf r},t)}{\partial t} = \dfrac{1}{\tau} \int_V G \bigl( |{\bf r} -
 {\bf r}_1| \bigr) F({\bf r}_1,t) \, dV_1 \\
 - \left(\dfrac{1}{\tau} + \sigma\right)F({\bf r},t) + q({\bf r},t).
 \label{eq:BH}
\end{multline}
where $\tau$ is the lifetime of excited atomic state with respect to spontaneous radiative decay;   
$\sigma$ is the rate of collisional quenching of excitation; $q$ is the source of excited atoms different 
from population by the absorption of the resonant photon (e.g., collisional excitation). 
The kernel $W$ is determined by the (normalized) emission spectral 
line shape $\varepsilon_{\omega}$ and the absorption coefficient $k_{\omega}$.
In homogeneous media, $W$ depends on the distance between 
the points of emission and absorption of the photon:
\begin{equation}
G(r) = - \dfrac{1}{4\pi r^2} \, \dfrac{dT(r)}{dr}, \quad
T(r) = \int_0^{\infty} \varepsilon_{\omega} 
\exp(-k_{\omega}r) \, d\omega.
\label{eq:GandT}
\end{equation}
The non-locality of the Biberman-Holstein radiative transfer demands special 
definition of the mean time $\bar t(r)$ needed for a photon to 
pass the distance $r$ from a point instant source $q({\bf r},t) = 
\delta({\bf r} - {\bf r}_0)\delta(t-t_0)$. The respective scalings for 
various line broadening mechanisms strongly deviate from 
the diffusion law (see \cite{bibvoyak:noneq,abkolis:vtp1987,vekl:jetp,bib:dan,kogan:iaea1968}). 
For Doppler and Lorentz line shapes, the results \cite{vekl:jetp} may be written in the unified form: 
$\bar t(\rho) \approx 1/[AT_{\text{as}}(\rho)]$
\cite{kogan:iaea1968}, where $\rho \equiv k(\omega_0)r$, 
and $T_{\text{as}}(\rho)$ is the asymptotics of the Holstein functional 
$T$ at $\rho \gg 1$.

Our numerical analysis of the Veklenko's Green function \cite{vekl:jetp} 
for various line shapes shows that the scaling defined by the equation (\ref{eq:frontdef})
gives good approximation for the time moment when $F(r,t)$
attains its maximum value at the distance $r$ from the source.
It is the analysis that enabled us to suggest using Eq.\ (\ref{eq:frontdef}) in 
automodel solutions in this and other radiative transfer problems.

For a short time, $\tau \ll t \ll t_{\text{fr}}(\rho)$ 
(or, equivalently, far in advance of front propagation coming 
at the distance $r$, $\rho \gg \rho_{\text{fr}}(t) \gg 1$), the asymptotics of the Green function 
\cite{vekl:jetp} for Doppler and Lorentz line shapes may be written in the form
\begin{equation}
F \approx t\,G(\rho),
\label{asymptBH1}
\end{equation}
which corresponds to the direct excitation of distant atoms by the photons 
in the far wings of the spectral line shape (i.e., by L\'evy flights). Asymptotic solutions in these 
particular cases suggested an extension of this law to a much broader class of non-local transport.

The Green function far behind the propagation front, 
$\rho \ll \rho_{\text{fr}}(t)$, or equivalently 
$t \gg t_{\text{fr}}(\rho) \gg \tau$, may be estimated 
assuming the local uniformity of the excitation due to 
the fast exchange of atoms in the core of the spectral line shape. 
The respective quasi-plateau solution in the 3D case takes the form:
\begin{equation}
F(r,t) \approx \dfrac{1}{\dfrac{4}{3} \pi (r_{\text{fr}}(t))^3}
\eta (r_{\text{fr}}(t) - r).
\label{eq:quasiplat}
\end{equation}
Comparison of Eq.\ (\ref{eq:quasiplat}) with numerical calculations of the exact 
Green function \cite{vekl:jetp} proves this asymptotic to give 
a good scaling for time dependence for various line shapes. 
However, the absolute values of the plateau in Eq.\ (\ref{eq:quasiplat}) and the asymptotics of the exact 
Green function may differ by a constant which amounts to a factor of unity 
for Doppler line shape and $\sim 200$ for Lorentz line shape. The large value of 
the constant may be explained by the longer precursor to the excitation 
front of Eq.\ (\ref{eq:quasiplat}) in the case of the longer PDF tail that, in turn, stems from 
a wider wings of the Lorentz line shape.

\begin{figure}
(a)˜\includegraphics[width=0.46\textwidth]{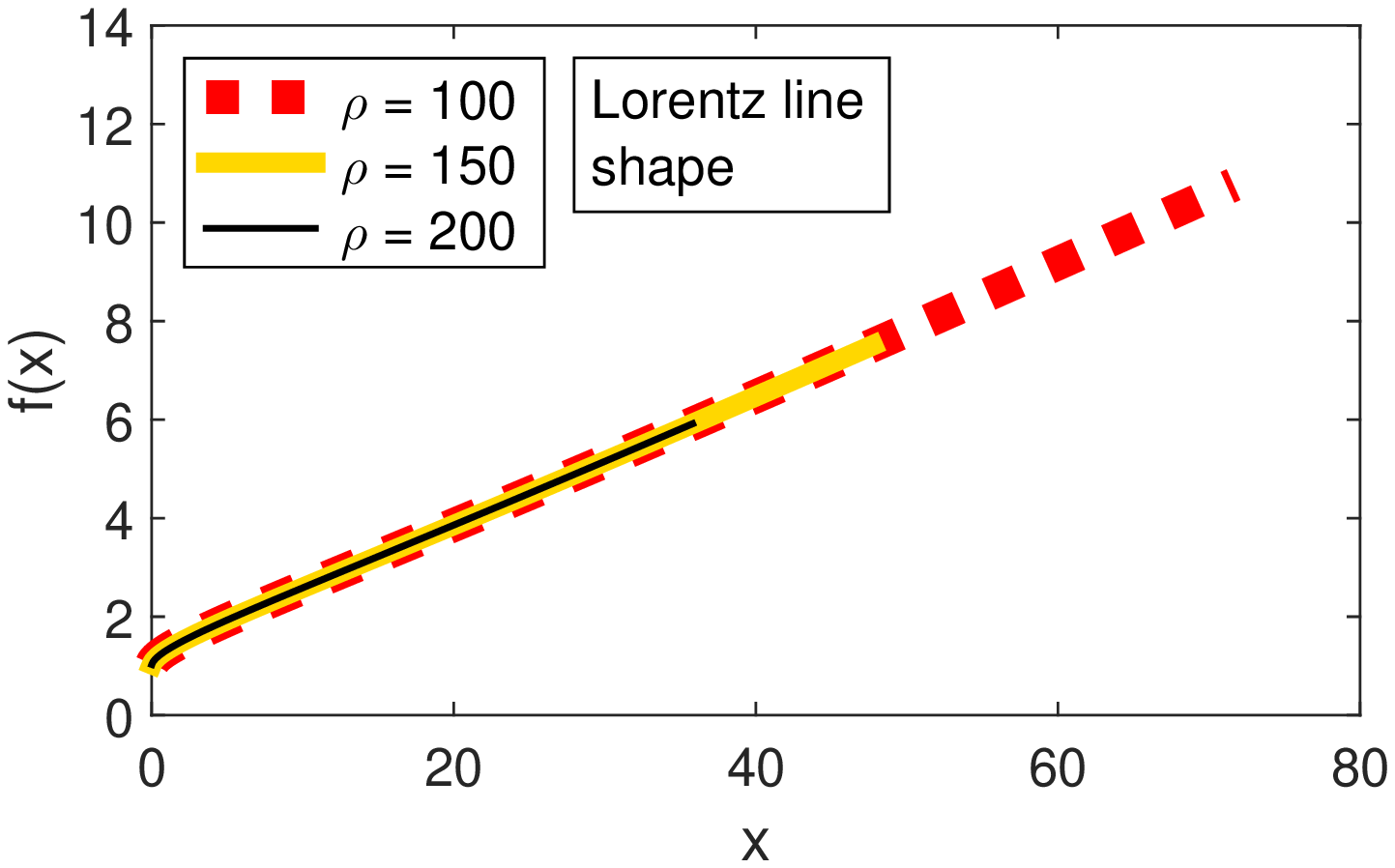} \\
(b)˜\includegraphics[width=0.46\textwidth]{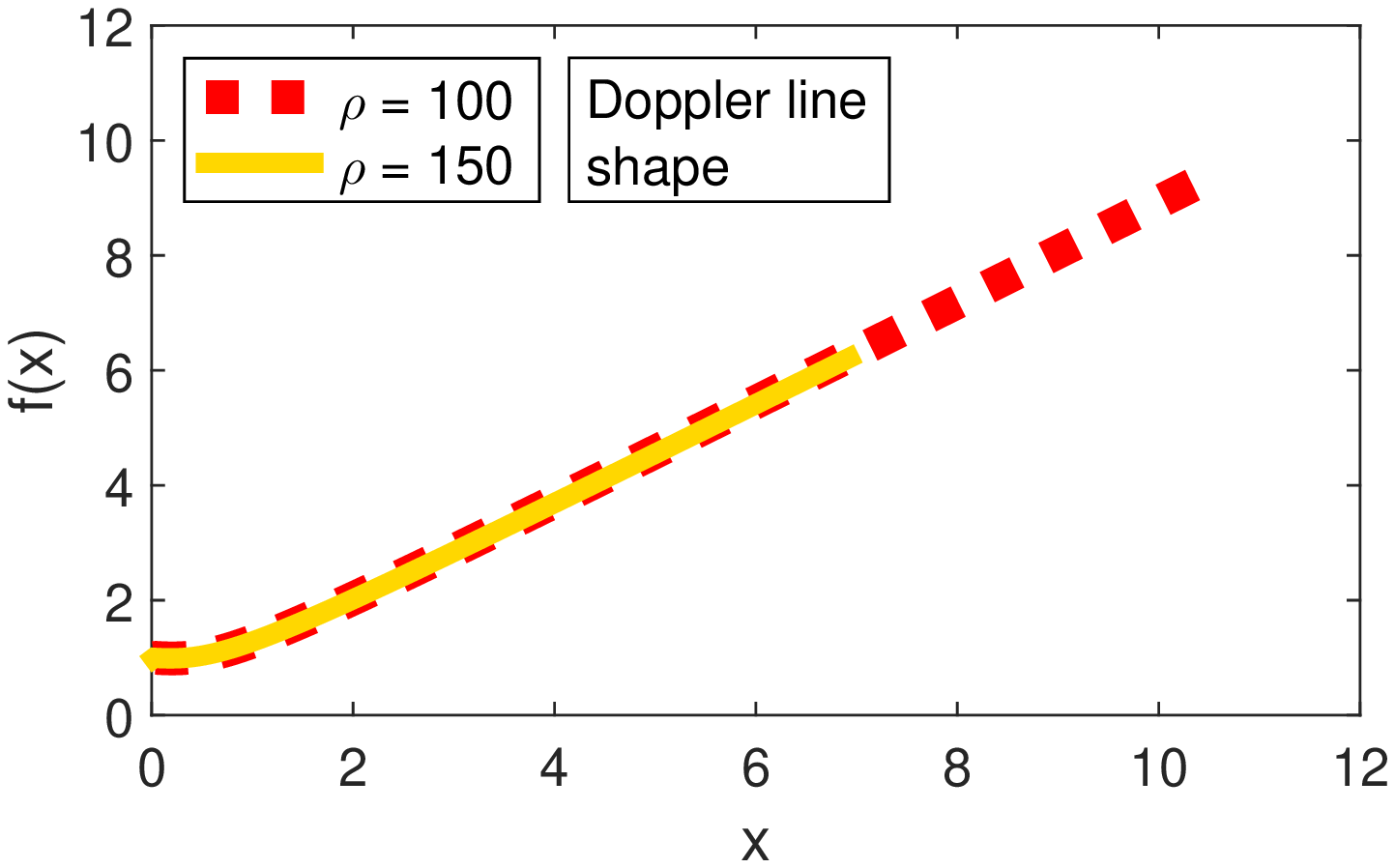} \\
(c)˜\includegraphics[width=0.46\textwidth]{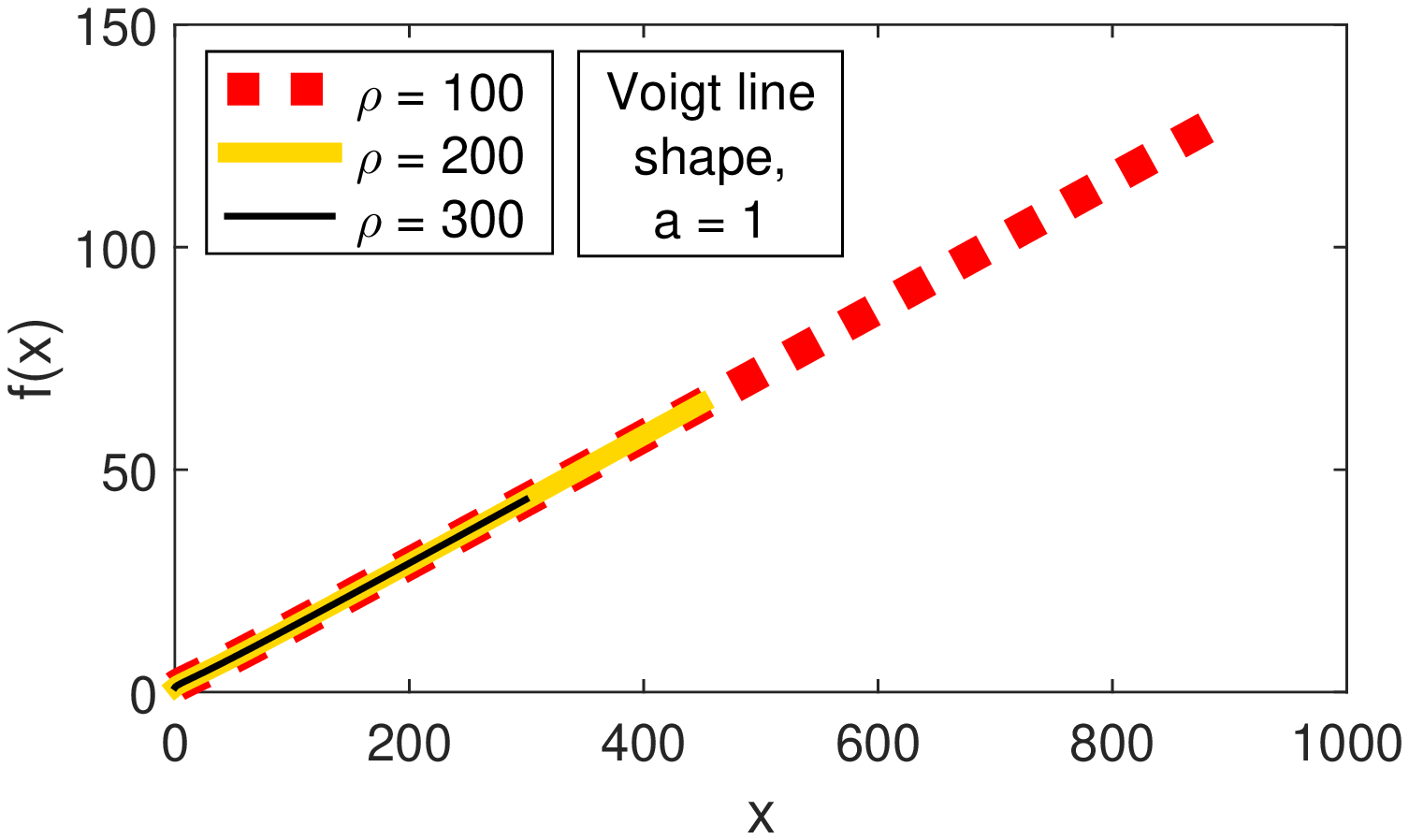} \\
(d)˜\includegraphics[width=0.46\textwidth]{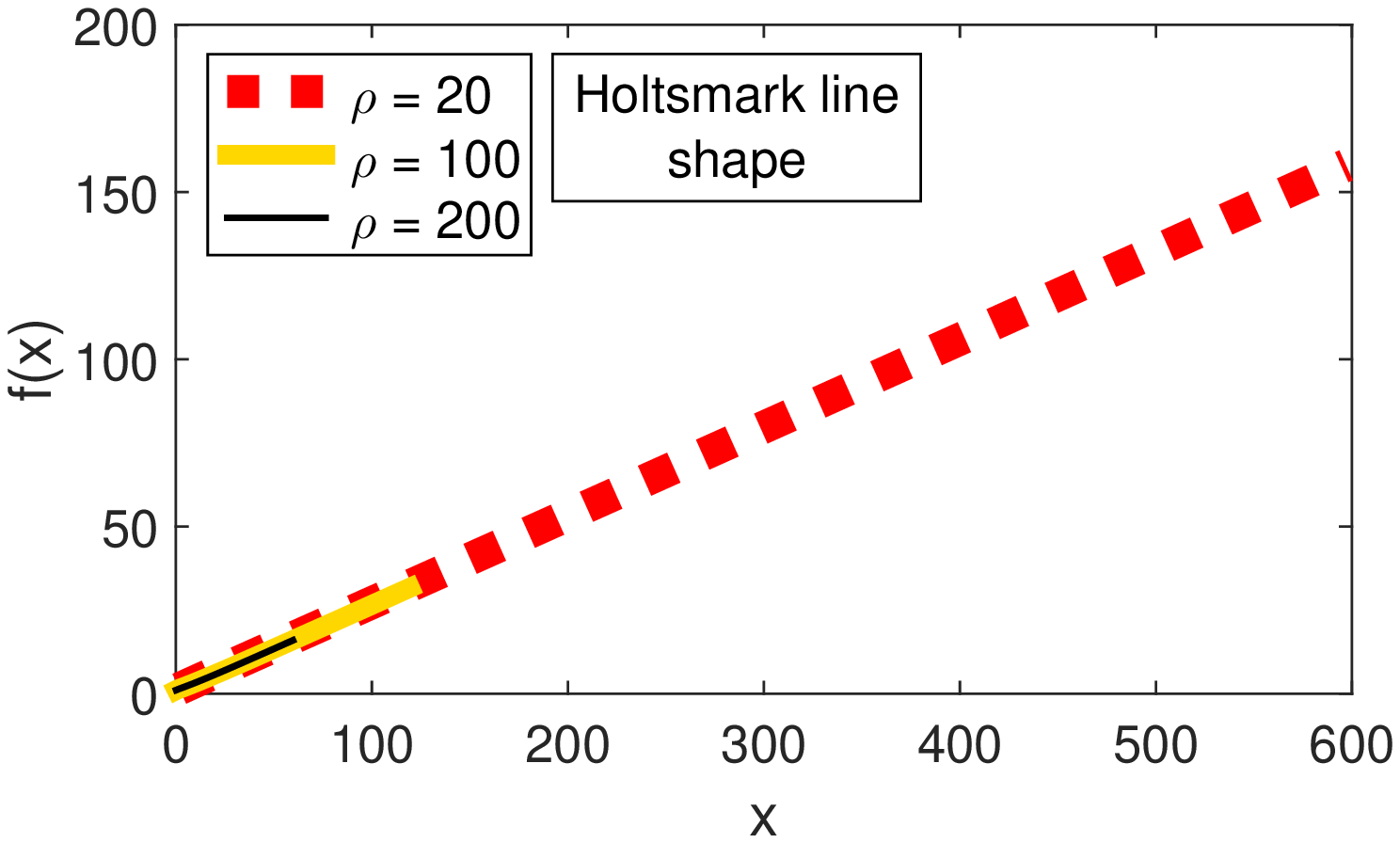}
\caption{\label{fig2} (Color online) Reconstruction of the argument of 
automodel function (\ref{eq:BHauto}), $x \equiv r_{\text{fr}}(t)/r$, from 
its comparison with the exact Green function \cite{vekl:jetp} for various 
spectral line shapes: (a) Lorentz; (b) Doppler; (c) Voigt for 
$a = \sqrt{\ln 2}\, \Delta\omega_{\text{Lorentz}}/
\Delta\omega_{\text{Doppler}}$, (d) Holtsmark.}
\end{figure}

The automodel solution is suggested in the form of Eq.\ (\ref{eq:interp2}) which in the 
3D case and for arbitrary space-time coordinates of the instant point source takes the form:
\begin{equation}
F_{\text{auto}}({\bf r},t; {\bf r}_0, t_0) = (t - t_0)
W \! \left( |{\bf r} - {\bf r}_0| \, 
f \left( \dfrac{r_{\text{fr}}(t - t_0)}{|{\bf r} - {\bf r}_0|} \right)
 \right).
\label{eq:BHauto}
\end{equation}
The results of the validation of automodel solution and the reconstruction of the function $g$ from 
comparison of the function (\ref{eq:BHauto}) with computations 
of the Green function \cite{vekl:jetp} 
for the Doppler, Lorentz, Voigt, and Holtsmark line shapes 
are shown in Fig.\ \ref{fig2}.

It is seen that the function (\ref{eq:BHauto}) is indeed an automodel Green function of Eq.\ (\ref{eq:BH}).

\section{\label{sec:concl}Conclusions}

The automodel solution of a wide class of non-stationary superdiffusive transport on 
a uniform background with a power-law decay, at large distances, of the step-length 
probability distribution function (PDF) is found. The success of identifying such solutions 
is based on the identification of the dominant role the long-free-path carriers (i.e. L\'evy flights) 
in all three scaling laws used to construct the automodel solution, namely, the scalings for 
the propagation front (i.e., relevant-to-superdiffusion average displacement) and asymptotic 
solutions far beyond and far in advance of the propagation front. The simplicity of the 
algorithm suggest the possibility to substantially broaden the applicability of the method.



%




\begin{acknowledgments}
The part of work related to the inverse problems solutions is supported by the 
Russian Foundation for Basic Research (grant RFBR-15-07-07850-a).
\end{acknowledgments}


\begin{thebibliography}{10}

\bibitem{shleza:ny}
\textit{L\'evy Flights and Related Topics in Physics}, edited by M. Shlesinger, G. M. Zaslavsky, and U. Frisch (Springer-Verlag, New York, 1995).

\bibitem{duspuch:bif}
A. A. Dubkov, B. Spagnolo, and V. V. Uchaikin, \textit{L\'evy flight superdiffusion: an introduction}, Int. J. Bifurcation Chaos \textbf{18}, 2649 (2008), DOI: 10.1142/S0218127408021877

\bibitem{mandelbrot}
B. B. Mandelbrot, \textit{The Fractal Geometry of Nature} (W.~H.~Freeman, New York, 1982).

\bibitem{bib:jetp}
L. M. Biberman, Zh. Eksper. Teor. Fiz. \textbf{17}, 416 (1947), [Sov. Phys. JETP \textbf{19}, 584 (1949)].

\bibitem{holstein:physrev}
T. Holstein, Phys. Rev. \textbf{72}, 1212 (1947).

\bibitem{bibvoyak:noneq}
L. M. Biberman, V.S. Vorob'ev, and I. T. Yakubov, \textit{Kinetics of Nonequilibrium Low Temperature Plasmas} (Consultants Bureau, New York, 1987).

\bibitem{abkolis:vtp1987}
V. A. Abramov, V. I. Kogan, and V. S. Lisitsa, in \textit{Reviews of Plasma Physics}, Vol. 12, edited by M. A. Leontovich and B. B. Kadomtsev (Consultants Bureau, New York, 1987), p. 151.

\bibitem{vekl:jetp}
B. A. Veklenko, Soviet Phys. JETP \textbf{9}, 138 (1959).

\bibitem{bib:dan}
L. M. Biberman, Dokl. Akad. Nauk SSSR \textbf{49}, 659 (1948).

\bibitem{kogan:iaea1968}
V. I. Kogan, in \textit{A Survey of Phenomena in Ionized Gases (Invited Papers), Proc. ICPIG’67} (IAEA, Vienna, 1968), p. 583 (in Russian).

\bibitem{radtransf:cambr}
W. Kalkofen, ed., \textit{Methods in Radiative Transfer} (Cambridge University Press, Cambridge, 1984), chapter 1.

\bibitem{napart:teplo}
A. P. Napartovich, Teplofiz. Vys. Temp. \textbf{9}, 26 (1971).

\bibitem{kuklissav:jetp}
A. B. Kukushkin, V. S. Lisitsa, and Yu. A. Saveliev, JETP Lett. \textbf{46}, 448 (1987).

\bibitem{rosliu:physfluids}
M. N. Rosenbluth, C. S. Liu, Phys. Fluids \textbf{19}, 815 (1976).

\bibitem{kuk:jetplett}
A. B. Kukushkin, JETP Lett. \textbf{56}, 487 (1992); \textit{Proc. 14th IAEA Conf. on Plasma Phys. Contr. Fusion}, Wuerzburg, 1992 (IAEA, Vienna, 1993), Vol. 2, p. 35; \textit{AIP Conf. Proc.} Vol. 299, \textit{Dense Z-pinches 3rd Int. Conf.}, London, 1993, edited by M. Haines and A. Knight (AIP Press, New York, 1994), p. 519.

\bibitem{tamor:fustech}
S. Tamor, Fusion Technol. \textbf{3}, 293 (1983); Nucl. Instr. and Meth. Phys. Res., A271, 37 (1988); ``A Simple Fast Routine for Computation of Energy Transport by Synchrotron Radiation in Tokamaks and Similar Geometries,'' SAI-023-81-189 LJ0LAPS-72, Science Applications (1981).

\bibitem{ivanov:radtransf}
V. V. Ivanov, \textit{Transfer of Radiation in Spectral Lines} (NBS Special Publication no 385) (US Govt Printing Office, Washington, DC, 1973).

\bibitem{mih:stellaratm}
D. Mihalas, \textit{Stellar Atmospheres} (W. H. Freeman \& Company, San Francisco, 1970) p. 399.

\bibitem{barberwier:nature}
P. Barthelemy, J. Bertolotti, and D. S. Wiersma, ``A L\'evy flight for light'', Nature (London) \textbf{453}, 495 (2008).

\bibitem{merchegue:pra}
N. Mercadier, M. Chevrollier, W. Guerin, and R. Kaiser, \textit{Microscopic characterization of L\'evy flights of light in atomic vapors}, Phys. Rev. A \textbf{87}, 063837 (2013).

\bibitem{uchai:physusp}
V. V. Uchaikin, \textit{Self-similar anomalous diffusion and L\'evy-stable laws}, Sov. Phys. Usp. \textbf{46}, 821--849 (2003).

\end{thebibliography}

\end{document}